\documentclass{ifacconf}

\usepackage{graphicx}      
\usepackage{natbib}        

\usepackage{mathrsfs}
\usepackage{amsfonts}
\usepackage{mathtools}
\usepackage{float}
\usepackage{subcaption}
\usepackage{graphicx}

\usepackage[colorinlistoftodos,prependcaption,textsize=footnotesize]{todonotes}

\begin{document}
\begin{frontmatter}
\title{Data-Driven Load-Current Sharing Control for Multi-Stack Fuel Cell System with Circulating Current Mitigation} 

\thanks[footnoteinfo]{This work was partially supported by the projects Supergen Energy Networks Hub (EP/S00078X/2) and Multi-energy Control of Cyber-Physical Urban Energy Systems (EP/T021969/1).}

\author[First]{Yiqiao Xu} 
\author[First]{Xiaoyu Guo} 
\author[Second]{Zhen Dong}
\author[First]{Zhengtao Ding} 
\author[First]{Alessandra Parisio}

\address[First]{University of Manchester, 
   Manchester, M13 9PL UK (e-mail: yiqiao.xu, xiaoyu.guo, zhengtao.ding, alessandra.parisio@ manchester.ac.uk).}
\address[Second]{University of Warwick, Coventry, CV4 7AL UK (e-mail: zhen.dong@warwick.ac.uk)}

\begin{abstract}  
The global trend toward renewable power generation has drawn great attention to hydrogen Fuel Cells (FCs), which have a wide variety of applications, from utility power stations to laptops. The Multi-stack Fuel Cell System (MFCS), which is an assembly of FC stacks, can be a remedy for obstacles in high-power applications. However, the output voltage of FC stacks varies dramatically under variable load conditions; hence, in order for MFCS to be efficiently operated and guarantee an appropriate load-current sharing among the FC stacks, advanced converter controllers for power conditioning need to be designed. An accurate circuit model is essential for controller design, which accounts for the fact that the parameters of some converter components may change due to aging and repetitive stress in long-term operations. Existing control frameworks and parametric and non-parametric system identification techniques do not consider the aforementioned challenges. Thus, this paper investigates the potential of a data-driven method that, without system identification, directly implements control on paralleled converters using raw data. Based on pre-collected input/output trajectories, a non-parametric representation of the overall circuit is produced for implementing predictive control. While approaching equal current sharing within the MFCS, the proposed method considers the minimization of load-following error and mitigation of circulating current between the converters. Simulation results verify the effectiveness of the proposed method.
\end{abstract}


\end{frontmatter}

\section{Introduction}
As attention to the environmental sustainability of urban districts and cities increases, hydrogen and Fuel Cells (FCs) are experiencing a growing interest as they offer the prospect of zero-emission power generation for a wide range of applications, in particular within the power and the automotive sectors. Though cases of successfully commercialized hydrogen FC systems continue to emerge, we are still in their developmental infancy and have many challenges to overcome before they can be deployed on a greater scale.

FC systems are subject to rapidly fluctuating loads either in stationary or automotive applications. As a consequence of its slow dynamics, an FC stack itself is not capable of responding to load transients as quickly as desired [\cite{Shakeri:2020}]. Thus, power electronic converters have been gradually accepted by the industry and academia as an effective interface between FC stacks and the DC bus. Since a single FC stack can hardly meet the diverse application needs, the Multi-stack Fuel Cell System (MFCS), which integrates two or more stacks, is a preferable option, especially for high-power applications~[\cite{Zhou:2022}]. Compared to single-stack systems, it possessed a longer lifetime and outperforms also in terms of efficiency and reliability~[\cite{Marx:2014}]. Different connections of converters can be adopted for different cases, where the parallel connection, i.e., one converter per FC stack, provides more degrees of freedom for MFCSs and allows individual control over each stack~[\cite{Somaiah:2016}]. 

Above all, the loads must be properly shared among all stacks in the MFCS, highlighting the need for a load-current (power) sharing strategy. The fuel efficiencies of several commonly used sharing strategies like equal sharing ~[\cite{Han:2018}], daisy chain~[\cite{Fernandez:2020}], and online identification~[\cite {Wang:2019}] are summarized in \cite{Zhou:2022}. It can be seen that equal sharing exhibits satisfactory efficiency throughout the whole power range, whereas the other two underperform either in the low-power range or in the high-power range. To achieve designated current (power) sharing, model-based control such as Model Predictive Control (MPC) is an attractive solution. With the surge of computing power, digitally controlled converters have been able to implement MPC with each time step executed in hundreds of or even tens of microseconds~[\cite{Oettmeier:2009}]. \cite{Long:2015} proposed a master-slave MPC method to achieve equal current sharing among paralleled converters in an MFCS. For interleaved boost converters, \cite{Xu:2022} applied an extension of MPC in which the inevitable parameter uncertainties were handled by an additional module based on the flatness theory. Aiming at performance consistency, \cite{Wang:2020} developed an adaptive current sharing method taking into account degradation of each stack; a simplified circuit model ignoring inductance/capacitance effects was used for controller design.

However, the aforementioned methods require an accurate circuit model and the achievable performance highly depends on the quality of system identification. Furthermore, power electronic converters may have to face the aging of semiconductor components during their service life, the control performance is likely to be affected by parameter changes [\cite{Alam:2014}]. In order to address the aforementioned shortcomings, we investigate the potential of data-driven methods in the filed of MFCS, which constitutes a major contribution of this paper. Different from the state-of-the-art methods relying on sequential system identification and control, this method directly adopts the raw data collected offline for load-current sharing control of MFCS. The proposed method is data-driven and model-free, and it is of great significance when circuit parameters are not exactly known. In the presence of unmatched Equivalent Series Resistances (ESRs), circulating current issues may arise among paralleled converters, leading to converter overstress and overheating [\cite{Zhang:2011}]. Thus, the load-current sharing strategy should be adapted from strictly equal sharing; to this purpose, the proposed method includes, as an additional objective, the mitigation of circulating currents, reaching a compromise between load-following, equal current sharing, and circulating current mitigation.

In this paper, we will focus on a subclass of FC named Proton Exchange Membrane (PEMFC). In the past few years, PEMFC has received great research interest due to its superior power density, near-zero gas emission, high efficiency, and fast start-up even at low temperatures~[\cite{Lv:2018}]. 
\begin{figure}
\begin{center}
    \includegraphics[width=8.4cm]{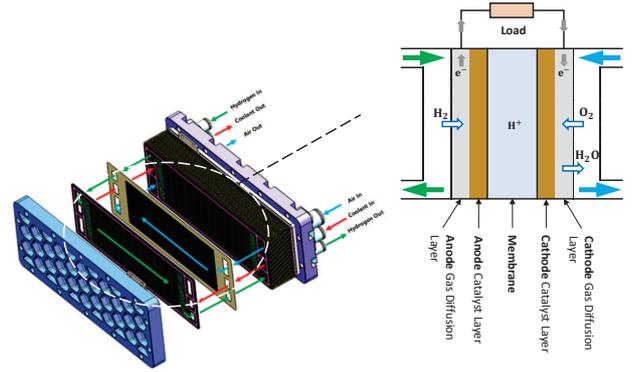}    
    \caption{Schematic diagram of a PEMFC stack and the partial cross-sectional view of a cell.} 
\end{center}
\end{figure}
The schematic overview of a hydrogen PEMFC stack is presented in Fig. 1, which stacks hundreds of cells in series to increase the output voltage. As the name suggests, the electrolyte of PEMFC is a polymeric membrane sandwiched by the anode and cathode. Each cell is a small-scale electrochemical system that converts the chemical energy of hydrogen, hydrocarbons, alcohols directly into electricity and heat. By pumping hydrogen into the anode and air into the cathode, the hydrogen and oxygen react across the membrane, resulting in a common current flowing through all cells [\cite{Wu:2016}].

\section{Overview of MFCS}
This section introduces the output characteristics of PEMFC stack and the equivalent model of MFCS. We would like to clarify that the proposed method is data-driven and model-free. The purpose of this section is to elaborate on the problem we study and to facilitate comparison with traditional model-based methods.

\subsection{I-V Output Characteristics}
The output voltage of a PEMFC stack, denoted by $V_{FC}$, can be expressed by the following formula:
\begin{align}\label{stack_model}
    V_{FC} = \left(E_{oc} - V_{act} - V_{ohm}\right) - E_{deg},
\end{align}
where $E_{oc}$, $V_{act}$, and $V_{ohm}$ correspond to the open circuit voltage (theoretical ideal voltage), activation polarization, and ohmic polarization, respectively. Models explicitly describing these polarization phenomena are available in~[\cite{Wang:2020}]. Furthermore, $E_{deg}$ is the voltage loss due to stack degradation and it can be characterized using a quadratic function of operation time.

\begin{figure}[htbp]
\begin{center}
    \includegraphics[width=8.4cm]{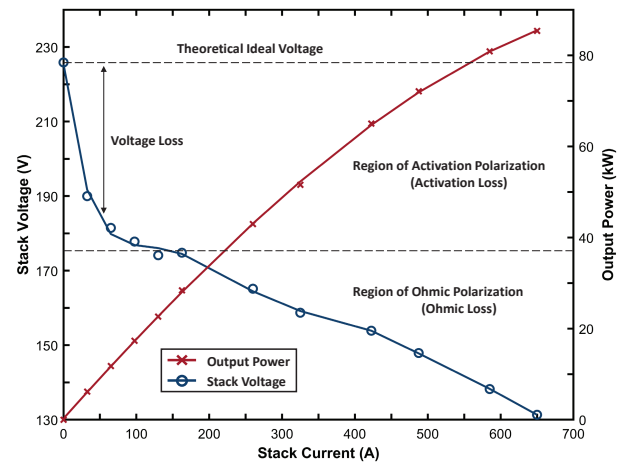}
    \caption{I-V polarization curve of an experimental PEMFC stack.}
\end{center}
\end{figure}
\begin{figure*}[htbp]
\begin{center}
    \includegraphics[width=14.0cm]{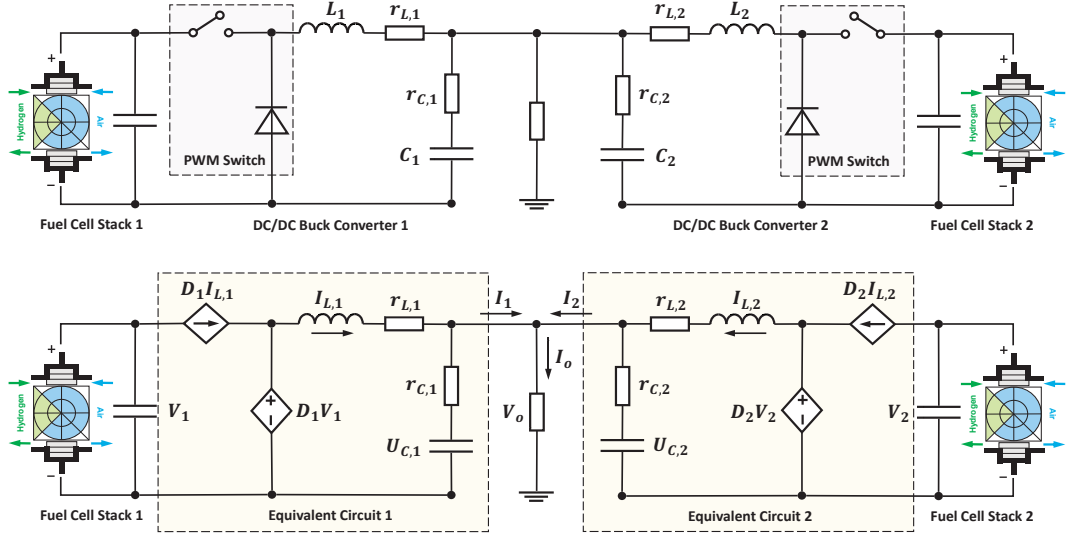}    
    \caption{Equivalent circuit model for the illustrative MFCS.}
    \label{fig:circuit}
\end{center}
\end{figure*}

With effective air/fuel flow control and thermal management, the polarization phenomena is dominated by the output characteristics of stacks. Fig. 2 illustrates an experimentally obtained I-V polarization curve of SEEEX 80 kW PEMFC stack. It can be seen that the stack voltage suffers from voltage drop mainly caused by ohmic loss and activation loss. The rapid voltage drop, prominent at low current densities, is governed by the activation polarization, while in the ohmic polarization region, the linear voltage drop can be attributed to the internal ohmic resistance. Another source of voltage loss called concentration polarization, which mainly occurs at extremely high current densities, is beyond the rated range and therefore omitted from (\ref{stack_model}) and Fig. 1.

\subsection{Equivalent Circuit Model}
Owing to its slow dynamics, a PEMFC stack itself is not capable of timely responding to load transients [\cite{Shakeri:2020}]. Therefore the stack voltage will be highly fluctuating at variable load conditions. In order to achieve performance compatible with load-following operation, power electronic converters are indispensable in an MFCS, reshaping the I-V output characteristics of MFCS [\cite{Somaiah:2016}].

Fig. 3 depicts an illustrative MFCS composed of two subsystems in parallel, each comprising an FC stack and a buck converter. The buck converters step the voltages down to adapt to a 120 V DC bus, which is common for DC microgrids~[\cite{Elsayed:2015}] and electric vehicle charging stations~[\cite{Rafi:2021}]. Each converter has an active switch and a passive switch, and its terminal voltage is adjusted by imposing an appropriate duty ratio on the active switch. For each subsystem, it follows that
\begin{align}
    V_{DC,i} &= D_iV_{FC,i},\\
    I_{FC,i} &= D_iI_{L,i}.
\end{align}
Here, $V_{FC,i}$ and $I_{FC,i}$ are respectively the stack voltage/current, whereas $V_{DC,i}$ and $I_{L,i}$ are respectively the converter terminal voltage and inductor current.

The equivalent circuit model in the bottom of Fig. 3 can be easily generalized to an arbitrary number of subsystems. Based on this, we can describe the dynamics of a general MFCS using the following equations based on Kirchhoff’s Voltage Law (KVL), considering all parasitic resistances:
\begin{align}
    &L_i\frac{dI_{L,i}}{dt} = -r_{L,i}I_{L,i} + D_iV_{FC,i} - V_o,\label{eq:KVL1}\\
    &r_{C,i}C_i\frac{dU_{C,i}}{dt} = -U_{C,i} + V_o,\label{eq:KVL2}\\
    &\forall i \in \{1,2,\dots,n\},\notag
\end{align}
where $L_i$ and $C_i$ are the output inductance and capacitance; $r_{L,i}$ and $r_{C,i}$ are respective inductor and capacitor ESRs; $U_{C,i}$ is the voltage across the output capacitor; $V_o$ is the DC bus voltage.

From Kirchhoff’s Current Law (KCL), we can derive the output current of the MFCS as
\begin{equation}\label{eq:KCL1}
\begin{split}
    I_{o} &= I_{1}+I_{2}+\cdots+I_{n}\\
    &=\sum_{i=1}^N\left(I_{L,i}+\frac{U_{C,i}}{r_{C,i}}\right) - V_o\sum_{i=1}^N\frac{1}{r_{C,i}},
\end{split}
\end{equation}
where 
\begin{equation}
    I_i = I_{L,i} - \frac{V_o-U_{C,i}}{r_{C,i}}
\end{equation}
has been used for deriving the second equality of (\ref{eq:KCL1}). 

Hybridization of FCs and Battery Energy Storage (BES) protect the MFCS from harmful transition while achieving higher fuel efficiency. Therefore an auxiliary BES is connected at the DC bus such that
\begin{align}\label{eq:KCL2}
    I_o + I_{b} = I_{l},
\end{align}
which describes the relationship of MFCS output current $I_o$, BES current $I_{b}$, and load current $I_{l}$. In most applications, the output load is unknown or time-varying, in which case an output current sensor is required to measure $I_{l}$ at the DC bus so that we can model them as an exogenous input. Note that the BES is treated as a controlled current source which improves transient responses.

According to (\ref{eq:KCL1}) and (\ref{eq:KCL2}), the bus voltage is a linear weighted sum of $I_{L,i}$, $U_{C,i}$, $I_{l}$:
\begin{equation}
    V_o = r_{C,tot}\left[\sum_{i=1}^N\left(I_{L,i}+\frac{U_{C,i}}{r_{C,i}}\right) - I_{l} + I_{b}\right],\label{eq:10}
\end{equation}
where $1/r_{C,tot} = \sum_{i=1}^n 1/r_{C,i}$.

Let $x\coloneqq[I_{L,1},\dots,I_{L,n},U_{C,1},\dots,U_{C,n}]^\top$ be the state vector, $u\coloneqq[D_1V_{FC,1},\dots,D_nV_{FC,n},I_{b}]^\top$ be the input vector, $\omega\coloneqq I_{l}$ be the disturbance. We can conclude
\begin{align}\label{eq:ss1}
\begin{split}
    \dot x &= \underbrace{\begin{bmatrix}
    {\rm diag}(\frac{-r_{L,i}}{L_i}) & O_{n,n}\\
    O_{n,n} & {\rm diag}(\frac{-1}{r_{C,i}C_i})
    \end{bmatrix}}_{\phi_1}x\\
    &\quad+ \underbrace{\begin{bmatrix}
    {\rm diag}(\frac{1}{L_i}) & O_{n,1}\\
    O_{n,n} & O_{n,1}
    \end{bmatrix}}_{\phi_2}u + \underbrace{\begin{bmatrix}
    {\rm col}(\frac{-1}{L_i})\\
    {\rm col}(\frac{1}{r_{C,i}C_i})
    \end{bmatrix}}_{\phi_3}V_o,
\end{split}
\end{align}
and similarly,
\begin{align}\label{eq:ss2}
    V_o = \underbrace{\begin{bmatrix} 1_n^\top r & {\rm row}(\frac{r_{C,tot}}{r_{C,i}})\end{bmatrix}}_{\phi_4}x + \underbrace{\begin{bmatrix} O_{1,n} & r_{C,tot}\end{bmatrix}}_{\phi_5}u + \underbrace{(-r_{C,tot})}_{\phi_6}w.
\end{align}

Let $y\coloneqq[I_{L,1},\dots,I_{L,n},V_o]$ be the output vector.
Combining (\ref{eq:ss1}) and (\ref{eq:ss2}), the time-averaged state-space model for the MFCS is given as follows
\begin{align}
&\dot x = Ax + Bu + E\omega,\label{eq:overal_ss1}\\
&y = Cx + Du + F\omega,\label{eq:overal_ss2}
\end{align}
where $A = \left(\phi_1+\phi_3\phi_4\right)$, $B=\left(\phi_2+\phi_3\phi_5\right)$, $E=\phi_3\phi_6$; $C=[I_n, O_{n,n};\phi_4]$, $D=[O_{n,n+1};\phi_5]$, $F=[O_{n,1};\phi_6]$.

\section{Load-Current Sharing Control}
\subsection{Circulating Current Issue}
Circulating currents are inevitable in such parallel structured generation systems, which comes as a major issue in the control of MFCSs. Mitigation of circulating currents should be considered to prevent overstress and overheating of converters.
\begin{figure}[htbp]
\begin{center}
    \includegraphics[width=7.5cm]{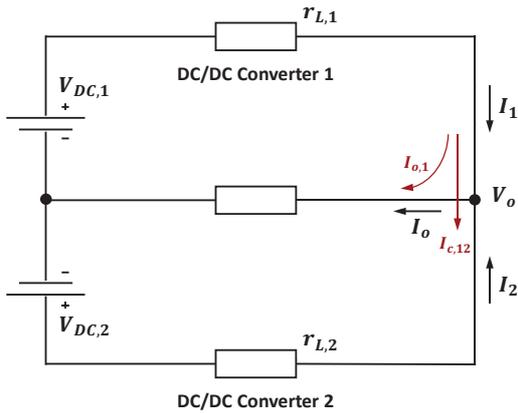}    
    \caption{Circulating current issue in a two-stack FC system.}
\end{center}
\end{figure}
In order to make the concept of circulating currents clear, Fig. 4 presents a reduced diagram for the equivalent circuit in Fig. 3. Note that they are equivalent at steady-state by invoking $C_i\dot U_{C,i}=0$. As illustrated above, the inductor current of each converter can be decomposed into a contributive term  $I_{c,i}$ and a harmful term $I_{c,ij}$, such that $I_{L,1} = I_{o,1} + I_{c,12}$ and $I_{L,2} = I_{o,2} + I_{c,21}$. 

By setting up the following KVL equations
\begin{align}
    V_{DC,1} = I_{L,1}r_{L,1} + I_{c,12}r_{L,2} + V_{DC,2},\\
    V_{DC,2} = I_{L,2}r_{L,2} + I_{c,21}r_{L,1} + V_{DC,1},\\
    V_{DC,1}-I_{L,1}r_{L,1} = V_{DC,2}-I_{L,2}r_{L,2},
\end{align}
we can obtain the circulating currents between two subsystems as
\begin{align}\label{eq:circulating_current}
\begin{split}
    I_{c,12} = -I_{c,21} &= \frac{V_{DC,1}-V_{DC,2}}{r_{L,1}+r_{L,2}}\\
    &= \frac{r_{L,1}I_{L,1}-r_{L,2}I_{L,2}}{r_{L,1}+r_{L,2}}.
\end{split}
\end{align}
Clearly, circulating currents appear whenever there exists an imbalance in terminal voltages of paralleled converters. Strictly equal current sharing is no longer suitable in the presence of unmatched ESRs, i.e. when $r_{L,1}\neq r_{L,2}$. Hence, we relax the goal of equal current sharing and additionally consider circulating current mitigation, which is achieved by including also the minimization of differences in converter terminal voltages.

\subsection{Optimization Problem Formulation}
We define the objectives of the proposed load-current sharing control of an MFCS as follow:
\begin{enumerate}
    \item Load-following: the subsystems and ancillary BES follow with the measured load current:
    $\vert \sum_{i=1}^n I_{L,i} + I_{b} - I_{l}\vert\rightarrow 0$;
    \item Equal current sharing and voltage regulation: the subsystems share the load current equally such that $\vert I_{L,i}-\bar I_L\vert\rightarrow 0$ ($\bar I_{L}\coloneqq\sum_{i=1}^n I_{L,i}/n$), and the DC bus voltage is maintained at its reference value, $\vert V_o-V_o^\star\vert\rightarrow 0$;
    \item Circulating current mitigation: $\vert V_{DC,i}-\bar V_{DC}\vert\rightarrow 0$ for $i\in\{1,\dots,n\}$, where $\bar V_{DC}\coloneqq\sum_{i=1}^n V_{DC,i}/n$.
\end{enumerate}

These objectives can be mathematically formulated by using the following quadratic cost functions:
\begin{equation}
\left\{
\begin{aligned}
    J_{1,k} &= \left(K_1y_k + K_2u_k - \omega_k\right)^\top Q_1\left(K_1y_k + K_2u_k - \omega_k\right),\\
    J_{2,k} &= \left(K_2y_k-r\right)^\top Q_2\left(K_2y_k-r\right),\\
    J_{3,k} &= \left(K_3u_k\right)^\top R\left(K_3u_k\right),
\end{aligned}
\right.
\end{equation}
where $K_1$, $K_2$, $K_3$ are linear mapping matrices that are straightforward to obtain; $Q_1$, $Q_2$, $R$ are performance indices that weight tracking errors and control efforts.

A constraint is imposed on inductor currents to protect the semiconductor switches, which is $0\leq I_{L,i}\leq  I_{L,i}^{\max}$. To safely supply downstream power devices, the DC bus voltage should lie within a narrow range. For this, we consider $V_o^{\min}\leq V_o\leq V_o^{\max}$. The duty ratio is limited to $[0.5,1]$, which implies $0.5D_iV_{FC,i}\leq D_iV_{FC,i}\leq V_{FC,i}$. In addition, there are bounds on the BES current, that is, $-I_b^{\max}\leq I_b\leq I_b^{\max}$. To summarize, the output vector and input vector are subject to definition domains, denoted by $\mathcal Y$ and $\mathcal U$.

\subsection{Data-Driven Predictive Control}
Since the proposed data-driven load-current sharing control inherits the basic structure of MPC, we first provide the formulation of a model-based MPC problem before focusing on the data-driven method. The MPC problem with a horizon $T_f$ is defined by
\begin{equation}\label{eq:MPC}
\begin{aligned}
    \min_{u,x,y} \quad & \sum_{k=0}^{T_f-1}J_k\coloneqq J_{1,k}+J_{2,k}+J_{3,k},\\
    \textrm{s.t.} \quad
    & x_{k+1} = A_dx_k+B_du_k+E_d\omega_k,\\ 
    & y_{k} = C_dx_k+D_du_k+F_d\omega_k,\\ 
    &\forall k\in\{0,\dots,T_f-1\},\\
    & x_0 = x(t),\\
    & y_k\in\mathcal Y,\ \forall k\in\{0,\dots,T_f-1\},\\
    & u_k\in\mathcal U,\ \forall k\in\{0,\dots,T_f-1\},
\end{aligned}
\end{equation}
where $A_d$, $B_d$, $C_d$, $E_d$, $F_d$ can be obtained by discretizing (\ref{eq:overal_ss1}) and (\ref{eq:overal_ss2}) at a specific sampling time. Their straightforward derivations are omitted here.

Given the challenges faced by model-based control, the idea of Data-Enabled Predictive Control (DeePC) [\cite{Coulson:2019}] has emerged in recent years. It has been adopted for the control of distribution networks [\cite{Bilgic:2022}], power plants [\cite{Huang:2021,Mahdavipour:2022}], and synchronous motor drives [\cite{Carlet:2022}]. Unlike MPC, DeePC is model-free, as it constructs a non-parametric representation of the system dynamics directly from raw data, which are a collection of input/output trajectories. 

To relate the future with the past, three Hankel matrices $\mathscr H(x^h)$, $\mathscr H(u^h)$, $\mathscr H(\omega^h)$ are built respective from historical trajectories $x^h$, $u^h$, $\omega^h$. Different lengths of historical trajectories can be selected by looking at the availability of data collection but each trajectory must be persistently exciting of order $L$, i.e., a Hankel matrix $\mathscr H_L(x)$
\begin{equation}
    \mathscr H_L(x)
    \coloneqq
    \begin{bmatrix}
    x_1 &x_2 &\cdots &x_{T-L+1}\\
    x_2 &x_3 &\cdots &x_{T-L+2}\\
    \vdots &\vdots &\ddots &\vdots\\
    x_L &x_{L+1} &\cdots &x_{T}
    \end{bmatrix}
\end{equation}
is of full row rank, where $T$ is the length of historical trajectories. This matrix can be partitioned into two sub-matrices, with one representing the inter-correlation among raw data for the past and the other for the future.

Therefore, under the persistency-of-excitation condition, there exists a unique vector $g$ such that the following equality holds for any fraction of system trajectories
\begin{equation}
\begin{bmatrix}
    \mathscr H_{T_p+T_f}(y^h)\\
    \mathscr H_{T_p+T_f}(u^h)\\
    \mathscr H_{T_p+T_f}(\omega^h)
\end{bmatrix}g=
\begin{bmatrix}
    y^p\\
    y^f\\ \hline
    u^p\\
    u^f\\ \hline
    \omega^p\\
    \hat\omega^f
\end{bmatrix}.
\end{equation}

At time instant $t$, $y^p\coloneqq[y(t-T_p)^\top,\dots,y(t-1)^\top]^\top$, $u^p\coloneqq[u(t-T_p)^\top,\dots,u(t-1)]^\top$, and $\omega^p\coloneqq[\omega(t-T_p),\dots,\omega(t-1)]^\top$ are the initial conditions in the length of $T_p$ for prediction, while $y^f\coloneqq[y(t)^\top,\dots,y(t+T_f-1)^\top]^\top$, $u^f\coloneqq[u(t),\dots,u(t+T_f-1)]^\top$, and $\omega^f\coloneqq[\omega(0),\dots,\omega(0)]^\top$ are future/current information in the length of $T_f$. This equality constraint serves a non-parametric representation and is being continuously updated as time horizon recedes.

Because of the implicit algebraic relations therein, DeePC minimizes over the objective function with four decision variables $y^f$, $u^f$, $g$, and $\sigma$. A robustification of DeePC against noisy measurement is adopted by introducing two penalty terms, $\lambda_g\Vert g\Vert^2$ and $\lambda_\sigma\Vert \sigma\Vert^2$, to the cost function. Then, problem (\ref{eq:MPC}) can be modified into
\begin{equation}\label{eq:21}
\begin{aligned}
    \min_{y^f,u^f,g,\sigma}\quad &\sum_{k=0}^{T_f-1}J_k+\lambda_g\Vert g\Vert^2+\lambda_\sigma\Vert\sigma\Vert^2,\\
    \textrm{s.t.} \quad&
    \begin{bmatrix}
        \mathscr H_{T_p+T_f}(y^h)\\
        \mathscr H_{T_p+T_f}(u^h)\\
        \mathscr H_{T_p+T_f}(\omega^h)
        \end{bmatrix}g=
        \begin{bmatrix}
        y^p\\
        y^f\\
        u^p\\
        u^f\\
        \omega^p\\
        \omega^f
    \end{bmatrix}+
    \begin{bmatrix}
    \sigma\\
    0\\
    0\\
    0\\
    0\\
    0
    \end{bmatrix},\\
    & y_0 = y(t),\\
    & y_k\in\mathcal Y,\ \forall k\in\{0,\dots,T_f-1\},\\
    & u_k\in\mathcal U,\ \forall k\in\{0,\dots,T_f-1\}.
\end{aligned}
\end{equation}
DeePC solves a finite horizon constrained optimization problem for each control interval $[\tau t,\tau(t+1))$. By optimizing a sequence of $u^f$ but applying only the first control action, the closed-loop system can be stabilized in a receding horizon way. 
 
\section{Simulation Results}
The performance of DeePC is validated in MATLAB and SIMULINK environments. To elucidate the load-current sharing control of MFCS, a dual-stack PEMFC system is considered for simulation. The two stacks are uniformly sized, each containing 330 cells, and therefore possess almost the same I-V output characteristics, except for the effect of degradation [\cite{Wang:2020}]. Different from the ideal case where paralleled converters are identical, the two converters slightly vary in their component parameters. The reference DC bus voltage is set as 120 V.

Historical trajectories including terminal voltages, BES current, load current, time-averaged inductor currents, and DC bus voltage are collected offline. Note that a necessary condition for persistency of excitation is $T\geq(m+1)(T_p+T_f)-1$, where $m$ is the output dimension. This can guide us to select a proper $T$ and the resulting Hankel matrices must have full row rank to ensure the efficacy of the collected historical trajectories. 

Regarding MPC for switched-mode converters, a popular option for the prediction horizon is unit-step or two steps~[\cite{Cortes:2010},~\cite{Saeed:2018},~\cite{Xu:2022}]. In this paper, we select a sampling time of 1 ms for DeePC/MPC and set $T_p=1$, $T_f=2$, and $T=30$, allowing adequate time for computation and subsequent implementation. Also, smaller time resolution can be adopted in practice, given the switching frequency of converters is typically over 20 kHz. Simulation results are given in Fig. 5 -- Fig. 6. 
\begin{figure}
     \centering
     \begin{subfigure}[htbp]{0.24\textwidth}
         \centering
         \includegraphics[width=\textwidth]{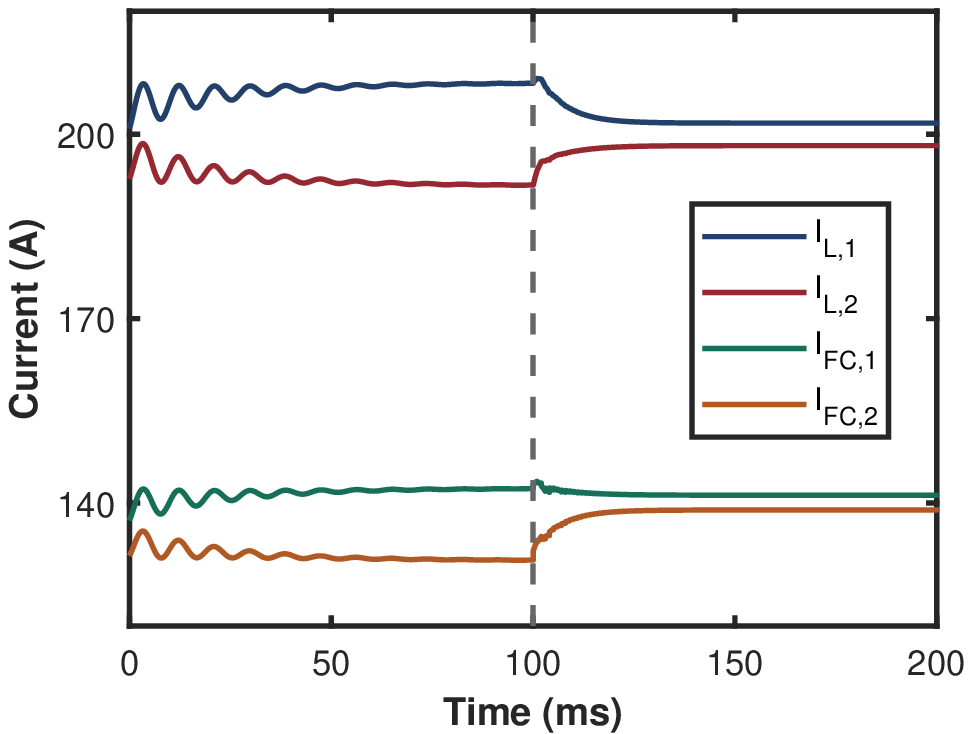}
         \caption{Inductor/stack currents}
     \end{subfigure}
     \begin{subfigure}[htbp]{0.24\textwidth}
         \centering
         \includegraphics[width=\textwidth]{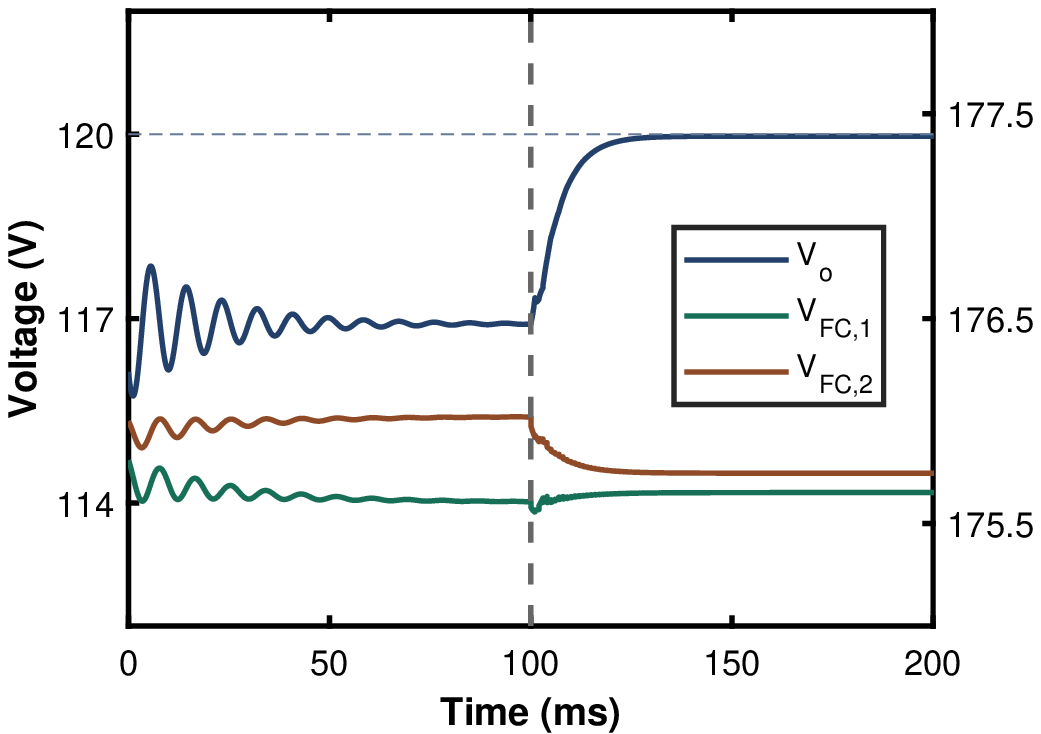}
         \caption{Bus (left)/stack voltages}
     \end{subfigure}
     \begin{subfigure}[htbp]{0.24\textwidth}
         \centering
         \includegraphics[width=\textwidth]{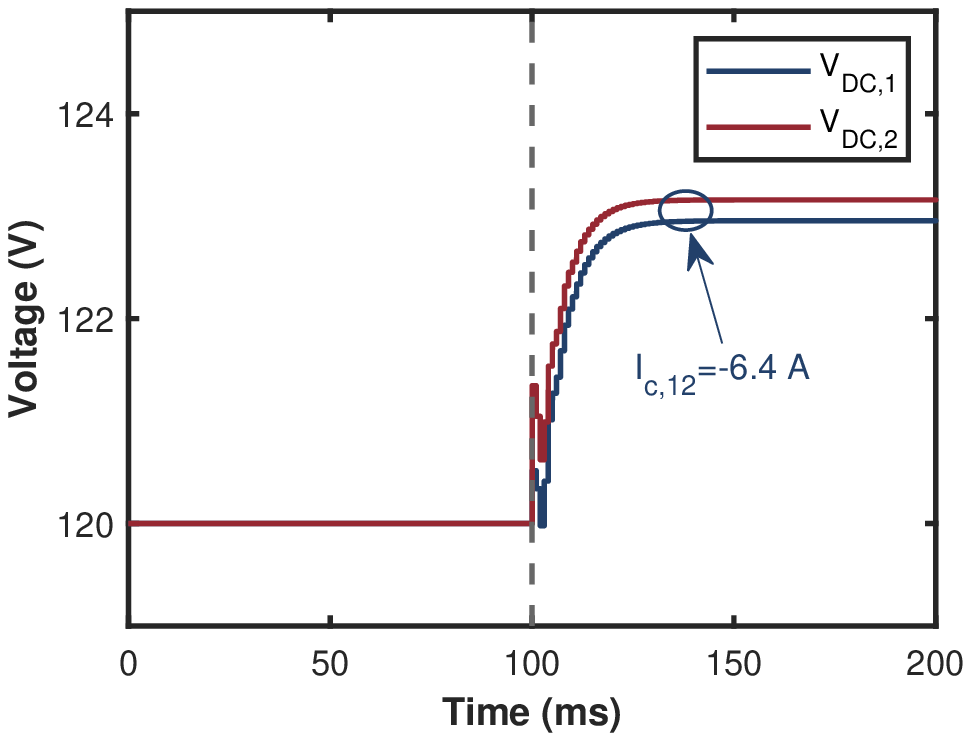}
         \caption{Converter terminal voltages}
     \end{subfigure}
     \begin{subfigure}[htbp]{0.24\textwidth}
         \centering
         \includegraphics[width=\textwidth]{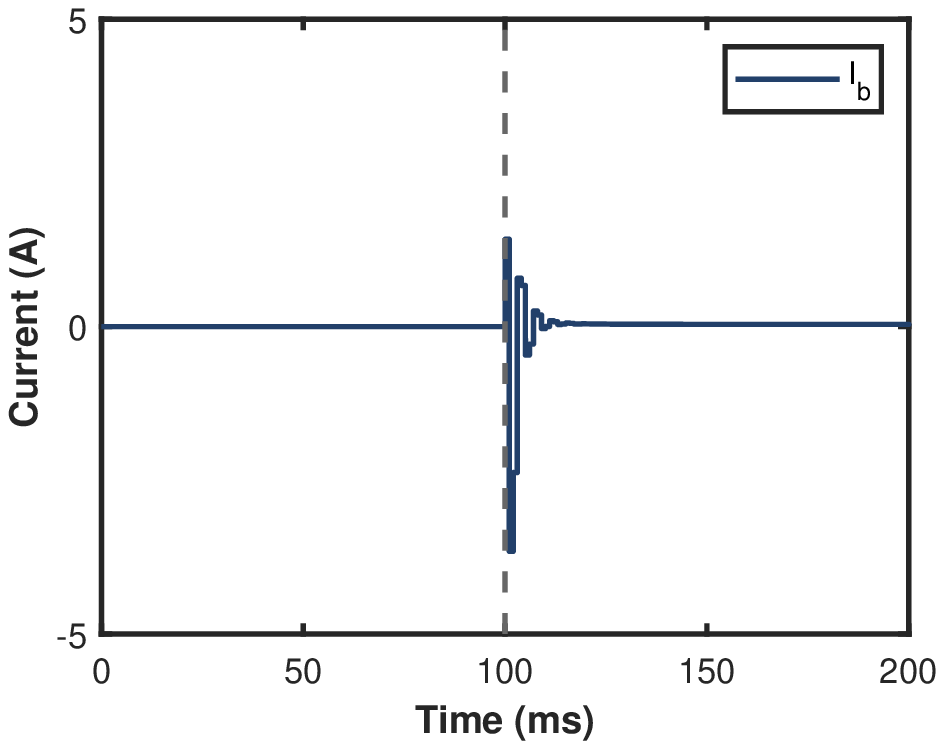}
         \caption{BES current}
     \end{subfigure}
     \caption{Responses of the MFCS with DeePC activated at 100 ms.}
\end{figure}

Firstly, the effectiveness of DeePC is revealed by activating DeePC at 100 ms. Constant load current of 400 A is adopted for both the activated phase (0--100 ms) and inactivated phase (100--200 ms). As shown in Fig. 5(c), the terminal voltages are simply kept at 120 V during the inactivated phase. As a consequence, the inductor currents and DC bus voltage gradually settle to constant values, with significant oscillations observed. However, they can be stabilized in less than 20 ms as long as DeePC is activated.

As shown in Fig. 5(b) and (d), the BES contributes mainly to transient responses and its output quickly recovers to 0 at steady-state. Due to a mismatch in inductor ESR of the two converters, the (time-averaged) inductor currents, as presented in Fig. 5(a), have to slightly deviate from strictly equal current sharing for the sake of circulating current mitigation. In this paper, $r_{L,1}=14.9$ $\mathrm {m}\Omega$ and $r_{L,2}=16.1$ $\mathrm {m}\Omega$ such that $I_{c,12}$ will be -8.42 A at steady-state, as previously analyzed in (\ref{eq:circulating_current}). As a result, the circulating current has been reduced by 24 \%. Then, DeePC is compared with MPC (assuming an accurate circuit model) in load-following operation, so as to demonstrate the engineering potential of DeePC. A high-frequency sine wave is introduced to mimic the fluctuating nature of the load current. The two control methods start with the same initial conditions. It can be seen from Fig. 6(b)--(c) that DeePC offers comparable results to classical MPC that assumes perfect knowledge of the model.

\begin{figure}
     \centering
     \begin{subfigure}[b]{0.48\textwidth}
         \centering
         \includegraphics[width=\textwidth]{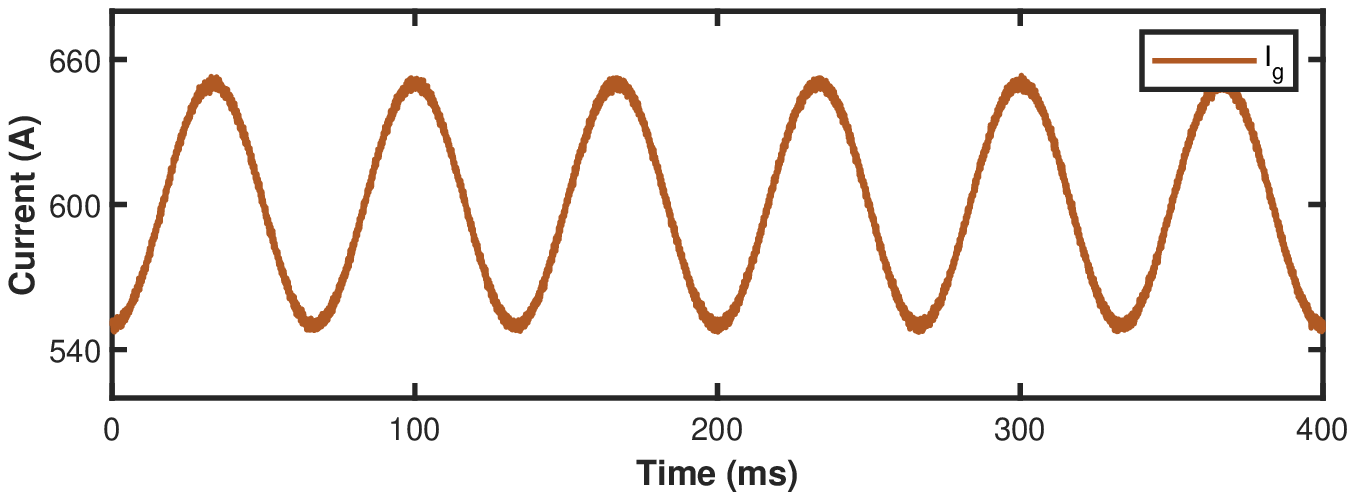}
         \caption{Load current}
     \end{subfigure}
     \hfill
     \begin{subfigure}[b]{0.48\textwidth}
         \centering
         \includegraphics[width=\textwidth]{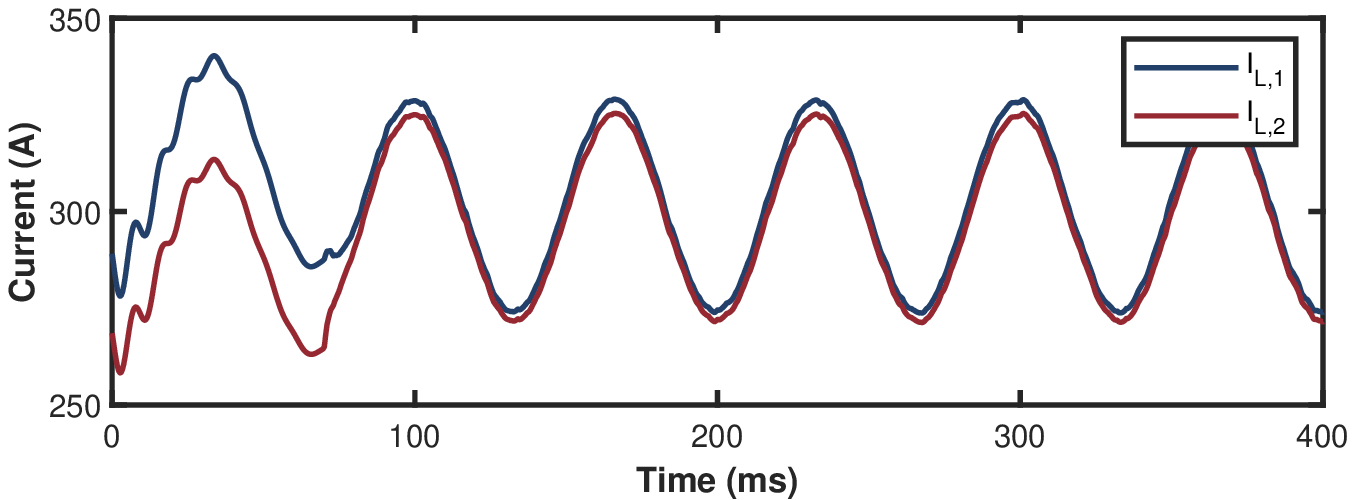}
         \caption{Inductor currents under DeePC}
     \end{subfigure}
     \begin{subfigure}[b]{0.48\textwidth}
         \centering
         \includegraphics[width=\textwidth]{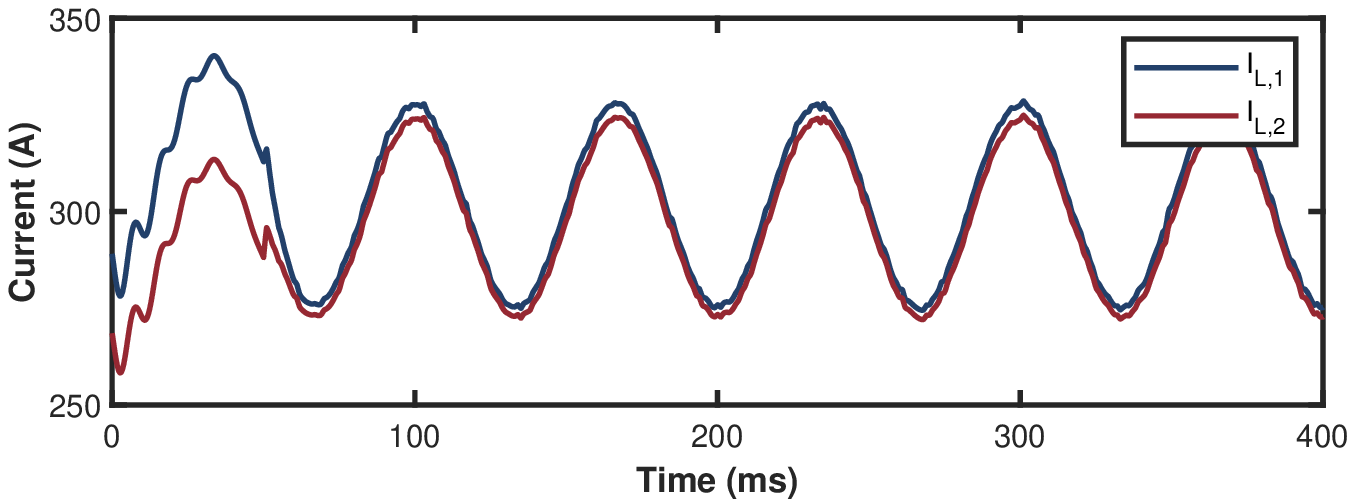}
         \caption{Inductor currents under MPC}
     \end{subfigure}
     \caption{Load-following performance of DeePC/MPC.}
\end{figure}
\section{Conclusion}
This paper explores the engineering potential of data-driven methods for load-current sharing control of MFCS. Paralleled converters have been controlled for power conditioning of PEMFC stacks. Simulation results demonstrate that the proposed method provides satisfactory control effects comparable to MPC which presumes precise model information. The proposed method is purely data-driven and model-free, suggesting great potentiality in the long-term operation of MFCSs whose circuit parameters may change unexpectedly. Our future directions include distributed data-driven control methods and their adaption for MFCS.

\begin{ack}
The authors would like to thank SEEEX Tech for the technical support and discussions on PEMFC and MFCS.
\end{ack}

\bibliography{ifacconf}

\end{document}